\documentclass[reprint, superscriptaddress, amsmath,amssymb, aps,floatfix]{revtex4-2}

\usepackage{notoccite}
\usepackage{natbib}
\usepackage{graphicx}
\usepackage{dcolumn}
\usepackage{bm}
\usepackage{color,graphicx,subfigure}
\usepackage{enumitem}
\usepackage{wrapfig}
\usepackage{soul}
\usepackage{amsmath}
\usepackage{titlesec}
\usepackage{multirow}
\usepackage{textcomp}
\usepackage{gensymb}
\usepackage[]{hyperref}
\hypersetup{colorlinks=true,linkcolor=blue,citecolor=blue,urlcolor=blue,pdfpagemode=UseNone}
\usepackage{xcolor}
\usepackage{etoolbox}
\usepackage{floatrow}
\usepackage{floatflt}
\usepackage{datetime}
\usepackage{graphicx}
\usepackage{dcolumn}
\usepackage{bm}
\usepackage{lineno}
\usepackage[normalem]{ulem}

\definecolor{eh}{rgb}{1, 0, 0}

\definecolor{ks}{rgb}{0.4, 0, 1}

\begin{document}
\preprint{APS/123-QED}

\title{Contrasting Spin Excitations in Octahedral and Square-Planar \textit{n}=8 Ruddlesden–Popper Nickelates}

\author{K.\,Scott}
\thanks{These authors contributed equally.}
\affiliation{\footnotesize \mbox{Department of Physics, Yale University, New Haven, Connecticut 06520, USA}}
\affiliation{\footnotesize \mbox{Energy Sciences Institute, Yale University, West Haven, Connecticut 06516, USA}}

\author{H.\, LaBollita}
\thanks{These authors contributed equally.}
\affiliation{\footnotesize Center for Computational Quantum Physics,
             Flatiron Institute,
             162 5th Avenue, New York, New York 10010, USA.}

\author{G.\,A.\,Pan}
\affiliation{\footnotesize Department of Physics, Harvard University, Cambridge, Massachusetts 02138, USA}

\author{X.\,Yang}
\affiliation{\footnotesize \mbox{Department of Physics, Yale University, New Haven, Connecticut 06520, USA}}
\affiliation{\footnotesize \mbox{Energy Sciences Institute, Yale University, West Haven, Connecticut 06516, USA}}

\author{A.\,Kar}
\affiliation{Donostia International Physics Center (DIPC), Paseo Manuel de Lardizábal, 20018, San Sebastián, Spain}

\author{C.\,Lim}
\affiliation{Donostia International Physics Center (DIPC), Paseo Manuel de Lardizábal, 20018, San Sebastián, Spain}

\author{A.\,Thorshov}
\affiliation{\footnotesize Department of Physics, University of California San Diego, La Jolla, California 92093, USA}

\author{D.\, Ferenc Segedin}
\affiliation{\footnotesize Department of Physics, Harvard University, Cambridge, Massachusetts 02138, USA}

\author{C.\,M.\,Brooks}
\affiliation{\footnotesize Department of Physics, Harvard University, Cambridge, Massachusetts 02138, USA}

\author{F.\,Yakhou-Harris}
\affiliation{\footnotesize European Synchrotron Radiation Facility, 71 Avenue des Martyrs, Grenoble F-38043, France}

\author{K.\,Kummer}
\affiliation{\footnotesize European Synchrotron Radiation Facility, 71 Avenue des Martyrs, Grenoble F-38043, France}

\author{N.\,B.\,Brookes}
\affiliation{\footnotesize European Synchrotron Radiation Facility, 71 Avenue des Martyrs, Grenoble F-38043, France}

\author{F.\,Boschini}
\affiliation{\footnotesize Advanced Laser Light Source, Institut National de la Recherche Scientifique, Varennes QC J3X 1P7 Canada}

\author{A.\,Frano}
\affiliation{\footnotesize Department of Physics, University of California San Diego, La Jolla, California 92093, USA}
\affiliation{\footnotesize Program in Materials Science and Engineering, University of California San Diego, La Jolla, California 92093, USA}

\author{J.\,A.\,Mundy}
\affiliation{\footnotesize Department of Physics, Harvard University, Cambridge, Massachusetts 02138, USA}

\author{E.\,H.\,da Silva Neto}
\affiliation{\footnotesize \mbox{Department of Physics, Yale University, New Haven, Connecticut 06520, USA}}
\affiliation{\footnotesize \mbox{Energy Sciences Institute, Yale University, West Haven, Connecticut 06516, USA}}
\affiliation{\footnotesize \mbox{Department of Applied Physics, Yale University, New Haven, Connecticut 06520, USA}}

\author{A.\,S\, Botana}
\affiliation{\footnotesize Department of Physics, Arizona State University, Tempe, Arizona 85287, USA}
\email{antia.botana@asu.edu}

\author{S.\, Blanco-Canosa}
\affiliation{Donostia International Physics Center (DIPC), Paseo Manuel de Lardizábal, 20018, San Sebastián, Spain}
\affiliation{\footnotesize IKERBASQUE, Basque Foundation for Science, 48013 Bilbao, Spain}
\email{sblanco@dipc.org}

\begin{abstract}

The discovery of superconductivity in reduced square-planar nickelates marked a major advance in identifying structural and electronic analogs to the high-$T_c$ cuprates. The more recent observation of superconductivity in parent Ruddlesden--Popper (RP) octahedral nickelates---with a clear difference in electron count with respect to cuprates---raises new questions about the nature of superconductivity across these related but distinct nickelate families. Here, we use Ni $L_3$-edge resonant inelastic x-ray scattering (RIXS) to probe the low-energy excitations in a representative compound of both families: the parent octahedral $n=8$ RP phase Nd$_9$Ni$_8$O$_{25}$ (p-RP), which is non-superconducting, and its reduced square-planar counterpart Nd$_9$Ni$_8$O$_{18}$ (r-RP), which exhibits superconducting correlations with a $T_c \approx 5$ K. The $n=8$ p-RP develops a spin-density-wave (SDW) ground state with ordering wave vector $q_{\mathrm{SDW}} = (1/4,\, 1/4)$, analogous to the bilayer RP, while the $n=8$ r-RP shows an elastic peak at $q^\star = (1/3,\, 0)$. Polarimetric RIXS shows that the p-RP exhibits low-energy spectra dominated by weakly dispersive paramagnons along the 0$\rightarrow\pi$ and $\pi \!\rightarrow\! \pi$ directions, whereas the r-RP with superconducting correlations displays dispersionless magnetic excitations. Our results comprehensively map out the spin excitations and reveal fundamental differences in the ground state between these two distinct structural families.
\end{abstract}

\maketitle

\begin{figure}
\includegraphics[width=\columnwidth]{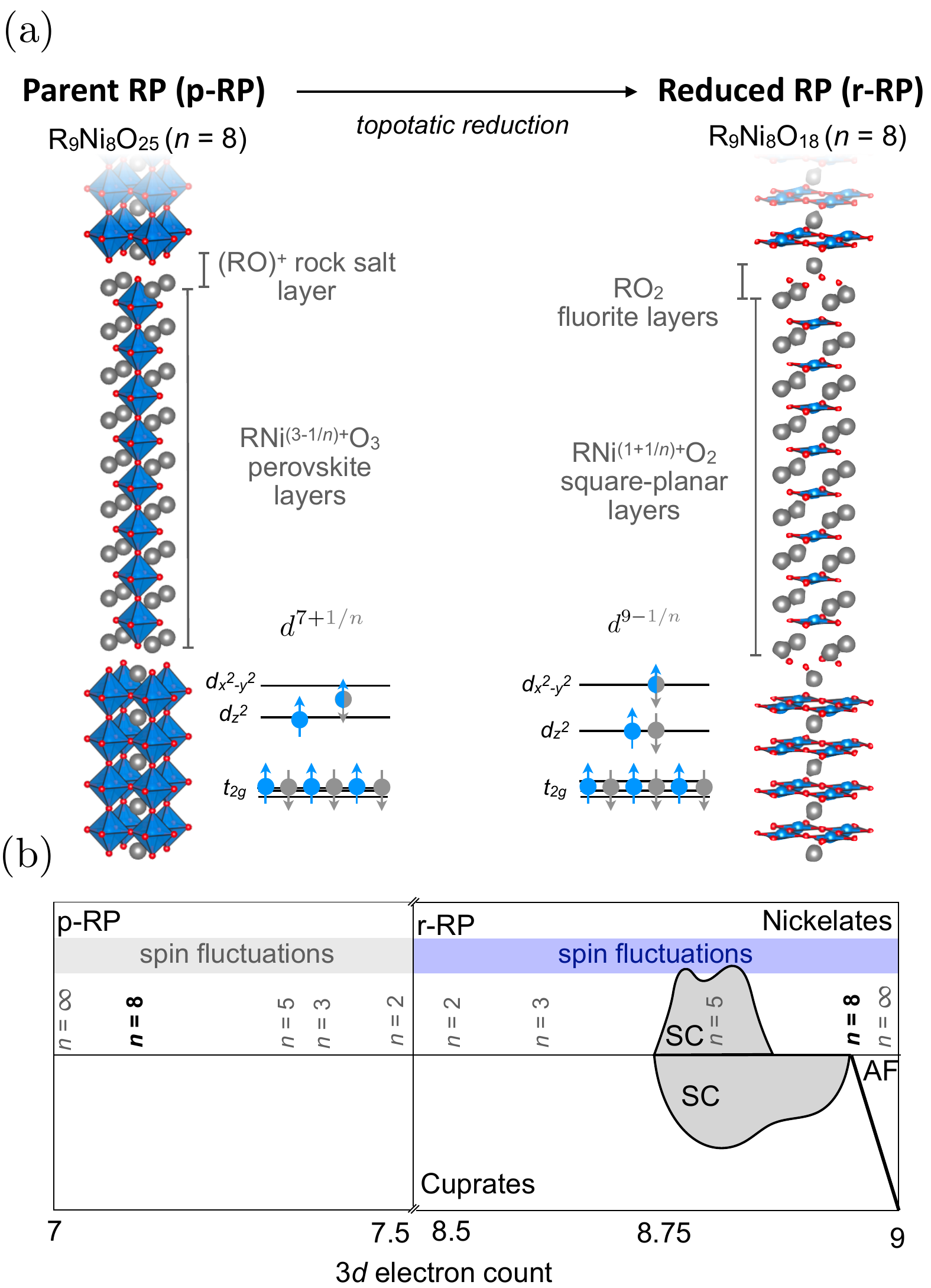}
\caption{\label{fig:fig1}(a) Qualitative overview of the structural and electronic properties of the $n=8$ parent (p-RP) and reduced Ruddlesden-Popper (r-RP) nickelates, where the reduced phase is obtained via topotactic reduction of the parent phases. (b)The schematic phase diagram as a function of $3d$ electron count denotes where the $n^{\mathrm{th}}$ line phase lies relative to well characterized electron fillings $d^{7}$, $d^{8}$, and $d^{9}$. A schematic of the hole-doped cuprate phase diagram denoting antiferromagnetic (AF) and superconducting (SC) phases is provided as reference. The 3\textit{d} electron count is defined as $d^{7+1/n}$ for p-RP and $d^{9-1/n}$ for r-RP, where \textit{n} is the number of Ni-O layers.}
\end{figure}

The search for cuprate-analog materials led to the discovery of superconductivity in infinite-layer nickelates RNiO$_2$ (R= rare-earth) with an analog $d^9$ filling and a square-planar coordination for the Ni ions \cite{Li_2019,osada2020, Osada2021_advmat,Puphal_2025}. Despite their obvious similarities to the cuprates, the infinite-layer nickelates have a much larger $p$-$d$ charge-transfer gap and a substantial hybridization between the Ni($3d$) and R($5d$) orbitals, unlike their copper counterparts \cite{Shen_2022,Botana2020,Hepting_2020,Bisogni_2016,Chen_2022, goodge2020, Ni1+isnotCu2+}.  In addition, the infinite-layer nickelates also display much lower T$_{c}$ ($\sim$ 15 K) \cite{Li_2019,osada2020, Osada2021_advmat}. Importantly, RNiO$_2$ compounds are the $n=\infty$ end members of the square-planar reduced RP series R$_{n+1}$Ni$_n$O$_{2n+2}$, containing $n$-NiO$_2$ planes and an average Ni filling $d^{9-1/n}$ \cite{LaBollita_2021, Puphal_2025}. This series is obtained via oxygen deintercalation of the parent R$_{n+1}$Ni$_n$O$_{3n+1}$ Ruddlesden-Popper phases \cite{greenblatt1997} (see Fig. \ref{fig:fig1}(a)). The finite Nd-based $n=4-7$ members of the reduced family have recently been shown to superconduct with similar physics and $T_\mathrm{c}$ to those of their infinite-layer counterparts \cite{Pan_2022,Pan2026}. More recently, signatures of superconductivity under pressure have been reported in the parent $n=2-3$ RP nickelates under pressure \cite{Sun_2023, wang2024bulk, hou2023emergence, zhang2023superconductivity, li2023signature,Schlömer_2024} and strain \cite{Ko_2025,Osada_2025}. The T$_c$'s of the parent RPs are much higher ($\sim$ 80 K for the bilayer, $\sim$ 30 K for the trilayer) despite their non-cuprate-like $d^{7+1/n}$ filling. Hence, two distinct but related families of superconducting nickelates are now available, spanning $3d$ electron fillings from $d^{9}$ to $d^7$ (see Fig.~\ref{fig:fig1}). 

Antiferromagnetic spin fluctuations play a relevant role in cuprates \cite{Dean_2014}. In the nickelates, resonant inelastic x-ray scattering (RIXS) measurements on the $n=3$ and $5$ parent (p-) and reduced (r-) thin film RPs have revealed non-dispersive low-energy magnetic excitations~\cite{Pan2026,Huisen_2025}, similar to the \textit{flattish} magnetic excitations in La$_{2-x}$Sr$_{x}$NiO$_{4}$~\cite{Fabbris_2017} and NdNiO$_{3}$~\cite{Lu_2018}. This contrasts with the trend observed in the $n=2$ p-RP single crystals and 
infinite-layer thin films that exhibit gapless magnetic excitations~\cite{Chen_2024,Lin_2021,Chen_2024a,Gupta_2025,Lu_2021,Yan_2025,Gao_2024}, akin to spin-$1/2$ antiferromagnet spin excitations on a cuprate-like square lattice. Therefore, the two nickelate families (p-RP and r-RP) seem to exhibit different magnetic excitation spectra that can be correlated with their structural and electronic differences from the available data at different $n$'s. 

 \begin{figure*}
\includegraphics[width=\textwidth]{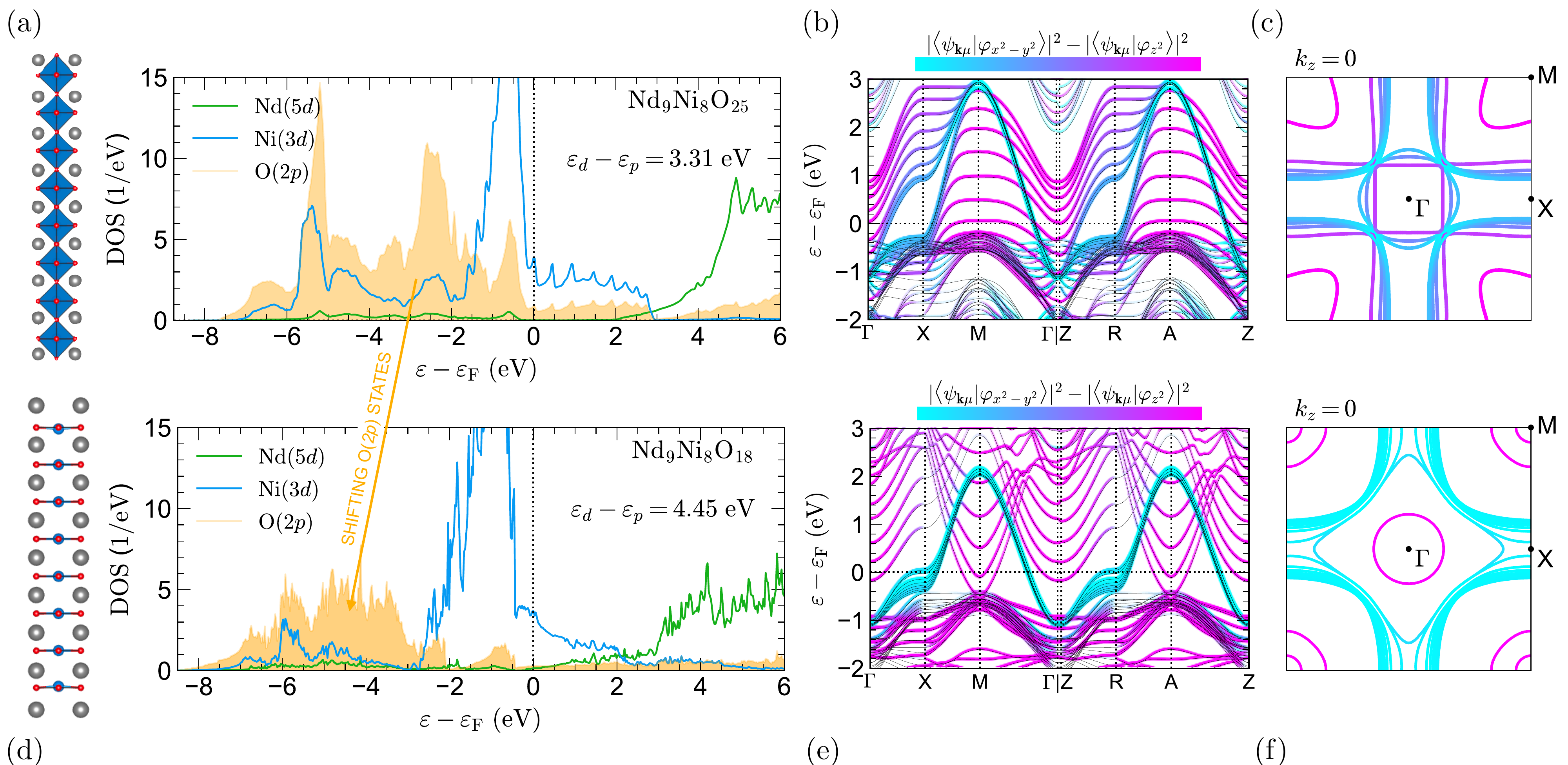}
\caption{\label{fig:fig2} DFT electronic structure and Fermi surface in the $k_{z}=0$ plane of the $n=8$ p-RP Nd$_{9}$Ni$_{8}$NiO$_{25}$ (a-c) and the r-RP Nd$_{9}$Ni$_{8}$O$_{18}$ (d-f).  Eight layer structural unit (left) and atom and orbital-resolved density of states: Nd($5d$) (green), Ni($3d$) (blue), and O($2p$) (orange) (right). DFT band structure along high-symmetry lines with Ni-$e_{g}$ character overlaid: Ni-$d_{x^{2}-y^{2}}$ (blue), Ni-$d_{z^{2}}$ (pink), and purple denoting the admixture of the two. Constant energy contour at $\varepsilon_{\mathrm{F}}$ in the $k_{z}=0$ plane with the Ni-$e_{g}$ orbital character overlaid using the same color scheme as the band structure.}
\end{figure*}

Here, we compare the ground state of the parent and reduced RP nickelates using the Nd-based $n=8$ compounds as a proxy by means of RIXS experiments at the Ni L$_3$-edge.  The $n=8$ r-RP (Nd$_{9}$Ni$_8$O$_{18}$), with a $d^{8.875}$ filling, falls directly in the underdoped region of the hole-doped cuprate phase diagram. The $n=8$ p-RP (Nd$_{9}$Ni$_8$O$_{25}$) has an average $d^{7.125}$ Ni filling, far from cuprate-like values (see Fig. \ref{fig:fig1}). We find that the p-RP exhibits an elastic peak with propagation vector $q_\mathrm{SDW}$=($\frac{1}{4}$, $\frac{1}{4}$), akin to that observed in the $n=2$ p-RP La$_3$Ni$_2$O$_7$ \cite{Chen-Donglai2024,Chen_2024a,Shi_2025,Zhao_2025}. Its low-energy inelastic spectra is characterized by a weakly dispersing optical magnon with frequency $\sim$55 meV along both high-symmetry directions, $0\rightarrow\pi$ and $\pi\rightarrow\pi$. In contrast, in the r-RP at a cuprate-like hole doping $x=0.125$, the long-range magnetic order is suppressed, and a new elastic peak develops with wavevector $q^*=(\frac{1}{3}, 0)$, potentially due to an oxygen-rich impurity phase that forms gradually through reoxidation processes \cite{Pan2026} as has been detected in the infinite-layer material \cite{Parzyck_2024}. The RIXS measurements reveal a higher energy ($\sim$60 meV) flat paramagnon that dominates the inelastic spectra over the lattice contribution.

High quality thin films of the parent $n=8$ RP Nd$_{9}$Ni$_8$O$_{25-\delta}$ and its reduced counterpart Nd$_{9}$Ni$_8$O$_{18-\delta}$  were grown by ozone-assisted molecular beam epitaxy (MBE) on NdGaO$_3$ substrates, as described in Refs. \cite{Pan_2022,Segedin_2023,Pan_2022a,Pan2026}. 
RIXS experiments, including polarization resolved measurements, were carried out at the ID32 beamline at the European Synchrotron Radiation Facility (ESRF) \cite{ID32,NBCO_pol,NCCO_pol,two_phonon_paper}, tuning the incident energy with $\sigma$ and $\pi$ polarization to the Ni-$L_3$ absorption edge (853 eV) at T=25 K. The combined energy resolution was $\Delta$E= 40 meV. To reveal the electronic structure of the $n=8$ p-RP and r-RP nickelates, Kohn-Sham density-functional theory (DFT) calculations were performed using an all-electron, full potential framework built on the augmented plane-wave plus local orbital (APW+lo) basis set as implemented in the WIEN2k code~\cite{Blaha_2020}. Complete details on the Methods are provided in the Supplementary Information \footnote{See Supplemental Material at [URL will be inserted by publisher] for additional details on RIXS data and analysis, polarimetric RIXS parameters, DFT and susceptibility calculations, and a comparison of magnetic excitations across RPNs, which includes Refs. \cite{Braicovich,NBCO_pol,NCCO_pol,two_phonon_paper,wien2k,vasp1,vasp2,vasp3,wannier90, wien2wannier,tprf,triqs}.}.

The crystal structures of the $n=8$ p-RP Nd$_{9}$Ni$_8$O$_{25}$ and of the corresponding r-RP Nd$_{9}$Ni$_8$O$_{18}$ are shown in Fig.~\ref{fig:fig1}(a). In the p-RP structure, eight consecutive NiO$_{6}$ perovskite layers are separated by a rocksalt spacer NdO$_{\mathrm{c}}$ slab, where O$_{\mathrm{c}}$ denotes cubically coordinated oxygen. Removing the apical oxygens (O$_{\mathrm{ap}}$) in the NiO$_{6}$ octahedra converts the rocksalt slab into a fluorite layer and isolates eight two-dimensional NiO$_{2}$ planes, giving rise to the r-RP phase that mirrors the La$_{1.875}$Ba$_{0.125}$CuO$_4$ and YBa$_2$Cu$_3$O$_{6.6}$ underdoped cuprates with hole doping \textit{p}= 0.125 \cite{Tranquada_1995,Blanco_2014,Frano_2020}.

The electronic structures of the octuple-layer parent and reduced RP phases are summarized in Figs.~\ref{fig:fig2}(a-c) and ~\ref{fig:fig2}(d-f), respectively. The band structure over a wide energy range for the $n=8$ p-RP is summarized  by the atomic-resolved density of states (DOS) in Fig.~\ref{fig:fig2}(a). A sizeable degree of Ni($3d$) and O($2p$) overlap can be observed, which translates into a charge-transfer energy $\Delta_{\mathrm{CT}}$= 3.3 eV. In this context, ligand hole 3$d^8\underline{L}$ physics is expected as indicated by the large amount of unoccupied O($2p$) density of states, similar to the bilayer RP \cite{dong2024}, La$_{2-x}$Sr$_x$NiO$_4$ \cite{Fabbris_2017} and the parent perovskite, NdNiO$_3$ \cite{Lu_2018,Bisogni_2016}. The low-energy electronic structure of the p-RP phase (see Fig.~\ref{fig:fig2}(b)) consists of eight dispersive Ni-$d_{x^{2}-y^{2}}$ (blue) bands hybridized with eight Ni-$d_{z^{2}}$ bands (pink), where the latter form molecular orbitals due to the hybridization with the apical oxygens and quantum confinement effects (see Fig.~\ref{fig:fig2}(a)). Figure~\ref{fig:fig2}(c) shows the p-RP Fermi surface in the $k_{z}=0$ plane with colors denoting the Ni-$e_{g}$ orbital character. In this plane, six hole-like sheets range from nearly pure Ni-$d_{x^{2}-y^{2}}$ (blue) to mixed $e_{g}$ (purple) character. A warped hole pocket at the zone corner arises from the flat M-point band and is entirely Ni-$d_{z^{2}}$ (pink). At $\Gamma$, three electron pockets appear: two circular Ni-$d_{x^{2}-y^{2}}$ (blue) pockets and one square pocket of mixed $e_{g}$ (purple) character. 

In contrast, the $n=8$ r-RP, exhibits a significant decrease in $p-d$ hybridization (see Fig.~\ref{fig:fig2}(d)), and hence a larger charge-transfer energy, $\Delta_{\mathrm{CT}}\sim 4.5$ eV. The low-energy electronic structure (Fig.~\ref{fig:fig2}(e)) consists of eight two-dimensional Ni-$d_{x^{2}-y^{2}}$ hole-like bands (blue) highly entangled with eight conduction states formed via Nd($5d$) and Ni-$d_{z^{2}}$ hybridized bands (pink). These rare-earth and Ni hybridized conduction states form the self-doping bands which effectively reduce the average Ni($3d$) filling from its nominal value of $d^{8.875}$. The Fermi surface shown in Fig.~\ref{fig:fig2}(f) results in an eight-sheeted (dominantly) Ni-$d_{x^{2}-y^{2}}$ bands with three additional self-doping electron-pockets (pink) around $\Gamma$ and M from the Ni-$d_{z^{2}}$-Nd hybridized conduction states. 

\begin{figure}
\includegraphics[width=1\textwidth]{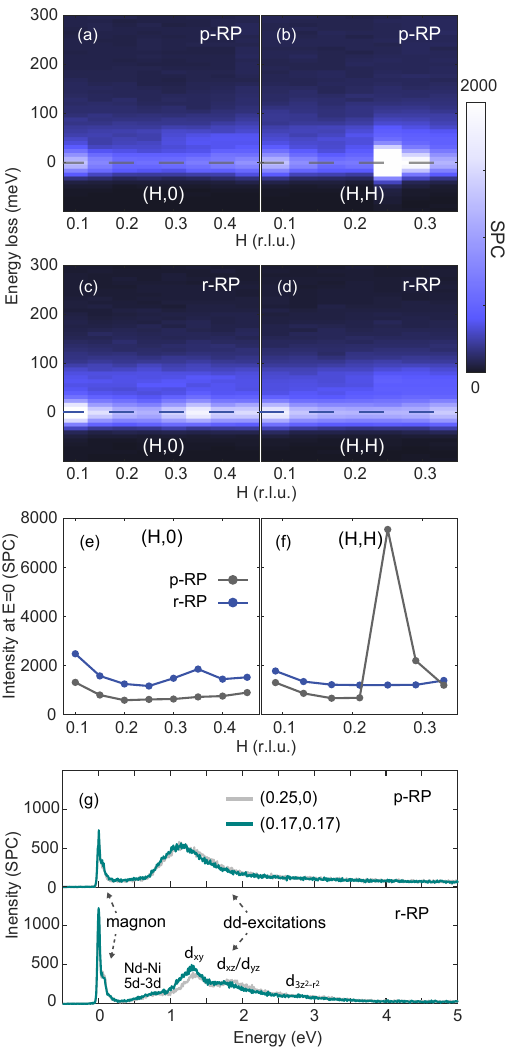}
\caption{\label{fig:fig3} (a-b) Low energy-loss RIXS maps along (H,\ 0) and (H,\ H) directions of the BZ of the p-RP and (c-d) of the r-RP. (e) Momentum profile of the elastic line at E$_\mathrm{loss}$=0 meV of the p- and r-RP along the (H,\ 0) direction. (f) Momentum dependence of the E$_\mathrm{loss}$=0 RIXS intensity of the p- and r-RPs along the (H,\ H) direction, highlighting the SDW at q=($\frac{1}{4}$, $\frac{1}{4}$). (g) RIXS spectra of the p- and r-RP, covering the 0-5 eV E$_\mathrm{loss}$ range. The orbital excitations are labeled in the r-RP together with signatures of the Nd-Ni hybridization.}
\end{figure}

With the DFT-level electronic structure of the octuple-layer nickelates established, Fig. \ref{fig:fig3} focuses on the RIXS results along the (0$\rightarrow\pi$ and $\pi\rightarrow\pi$) paths of the Brillouin Zone (BZ). The elastic signal (E$_\mathrm{loss}$=0) of the p-RP remains nearly constant with the momentum transfer along the (H,\ 0) direction (Fig. \ref{fig:fig3}(a)) but develops a sizable intensity with $\pi$-polarized light at (H,\ H)=($\frac{1}{4}$, $\frac{1}{4}$), diagonal to the Ni-O bond (Fig. \ref{fig:fig3}(b)), suggesting a magnetic origin for the static modulation (Supplementary Information Fig. S1), in agreement with the static magnetic order reported in 3D perovskites, \cite{Scagnoli_2008,Ren_2025,Frano_2013,Song_2023}, and in bilayer p-RP nickelates, where SDW formation has been widely discussed \cite{Chen_2024a,Fukamachi_2001,Liu_2022,Zhao_2025,Zhang_2020}. The calculated bare susceptibility reveals a peak near ($\frac{1}{4}$, $\frac{1}{4}$) for the p-RP (see Supplementary Information Fig. S5), consistent with the propagation vector of the SDW observed in the experiment. In the r-RP film, this peak is absent (Fig. \ref{fig:fig3}(c-d) and (f)), pointing to the critical role of the O$_\mathrm{ap}$-$p_z$ orbitals in the development of long-range magnetic order in the p-RPs. On the other hand, a weak signal can be discerned at (H,\ H)=($\frac{1}{3}$, 0) along the 0$\rightarrow\pi$ direction in the $\sigma$-channel in the r-RP (Fig. \ref{fig:fig3}(e)). Although further diffraction and microscopy measurements are required to elucidate its microscopic nature, the formation of oxygen superstructures due to oxygen-rich impurity phases caused by reoxidation processes \cite{Pan2026} is a likely cause for the presence of this ($\frac{1}{3}$, 0) peak \cite{Parzyck_2024,Pelliciari_2023,Tam_2023}.

The inelastic part of the RIXS spectra of both p- and r-RP is parsed in Fig. \ref{fig:fig3}(g). In the inelastic energy range between 0.6$<$E$_\mathrm{loss}<5$ eV, there are four major excitations that arise from the Nd-Ni hybridization ($\sim$0.7 eV) and from the optically forbidden \textit{dd}-excitations. In agreement with the DFT calculations, signatures of the Nd-Ni hybridization are suppressed in the p-RP and the \textit{dd} excitations appear as a broad single excitation (spectral width $\sim$1.5 eV), characteristic of a 3\textit{d}$^8$$\underline{\textit{L}}$ electronic configuration \cite{Huisen_2025,Fabbris_2017,Fursich_2019}. In the r-RP, the \textit{d}$_{z^2}$ orbital excitation can be observed at E$_\mathrm{loss}$$\sim$2.7 eV, the \textit{d}$_{xz/yz}$ at $\sim$1.8 eV, and the \textit{d}$_{xy}$ at $\sim$1.2 eV \cite{Gao_2024}. Between 20 meV$<$E$_\mathrm{loss}<$300 meV, the E$_\mathrm{loss}$ maps of both p- and r-RPs display a broad peak centered at $\sim$70 meV followed by a linear increase of the background, whose spectral weight results from a tail of the Nd-Ni hybridization and the higher energy \textit{dd} excitations, Supplementary Information Fig. S2. The overall energy scale of the excitations aligns with that observed in the \textit{n}=3 and \textit{n}=5 p- and r-RP nickelates and their frequencies are consistent with both a magnon and a phonon mode \cite{Huisen_2025,Pan2026}. 

\begin{figure}
\includegraphics[width=1.0\textwidth]{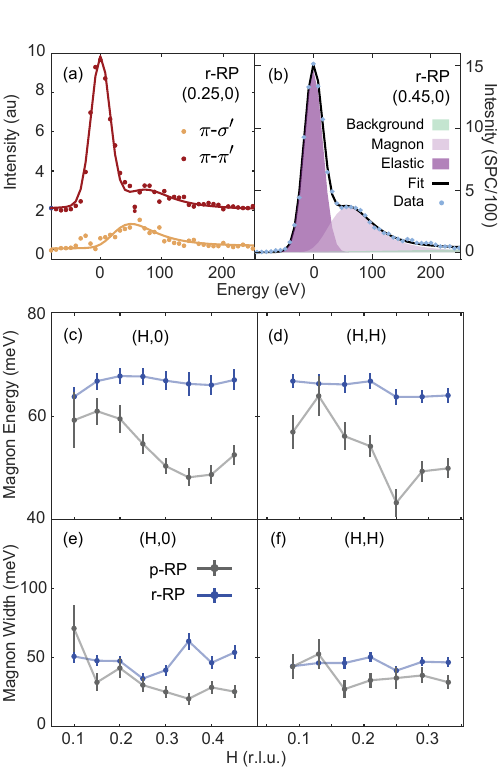}
\caption{\label{fig:fig4} (a) pol-RIXS data at q=(0.25,0) r.l.u. in the r-RP sample, showing the non-cross polarized channel ($\pi-\pi^\prime$) in red and the cross polarized channel ($\pi-\sigma^\prime$) in yellow. (b) Example fitting at q=(0.45,0) r.l.u. in the r-RP sample. An example fitting of a p-RP spectra, which is the same model aside from the background, can be found in Supplementary Figure S2. (c-d) Momentum dependence of the spin excitations of p- and r-RP along the (c) (H,\ 0)  and (d) (H,H) directions. (e-f) Damping of p- and r-RP along the (e) (H,\ 0)  and (f) (H,H) directions. The 95\% confidence intervals are plotting for each data point.}
\end{figure}

To disentangle overlapping magnetic (cross-polarized) and phononic (non-cross-polarized) contributions to the low-energy RIXS spectra, we have taken polarization-resolved RIXS (pol-RIXS) measurements at \textit{q}= ($\frac{1}{4}$, 0) r.l.u. for the r-RP, Fig. \ref{fig:fig4}a. The cross-polarized channel, defined as $\pi - \sigma^\prime$, clearly identifies a magnon excitation centered at $\sim$50 meV, in agreement with the energy scale of the magnetic excitations \textit{n}=3, 5 and $\infty$ nickelates \cite{Lu_2021,Gao_2024}. The non-cross-polarized channel ($\pi - \pi^\prime$) reveals an elastic signal and a phonon mode centered around 80 meV. Since the magnon signal dominates the intensity of the inelastic spectra in this energy region, and its energy position is well-centered in the momentum space, we approximate the convolved inelastic feature as a damped magnetic excitation, Fig. \ref{fig:fig4}b; see Supplemental Note III  for more details on the fitting procedure. The resulting fitting of the spin excitations in the p-RP reveals a weak wave-like dispersive excitation ($\sim$50-60 meV) along both the (H,\ 0) and (H,\ H) directions, Fig. \ref{fig:fig4}(c-d). On the other hand, the r-RP magnetic spectrum is characterized by a higher energy dispersiveless excitation ($\sim$ 65 meV), Fig. \ref{fig:fig4}(a-b). Despite the different chemical structure, dimensionality, and magnetic ground state, we find no substantial difference in the damping, behaving nearly constant across momentum, Fig. \ref{fig:fig4}(e-f).

The RIXS data presented here for the \textit{n}=8 p- and r-RPs align with the paramagnon frequencies and linewidths of the  parent and reduced \textit{n}=3 and 5 RP thin films \cite{Huisen_2025,Pan2026}. However, the \textit{n}=8 p-RP differs from the dispersive spin excitations in the \textit{n}=2 p-RP bulk samples~\cite{Chen-Donglai2024}, in spite of seemingly having the same magnetic order. Importantly, the strength of the paramagnon excitations in the \textit{n}=8 r-RP indicates a slight increase of the in-plane exchange interaction, \textit{J}$_\parallel$, provided as the interlayer exchange interaction (\textit{J}$_\bot$) weakens upon removal of the apical oxygens. The dominance of \textit{J}$_\parallel$ would favor superconducting theories that render a dominant role to the \textit{d}$_{x^2-y^2}$ orbital in this square-planar reduced RP phase \cite{Wu_2020,DiCataldo_2024}.

The electronic and magnetic ground state revealed in the $n=8$ r-RP nickelate with superconducting correlations, inevitably calls for a direct comparison with the cuprates at the same hole-doping level \textit{x}=0.125. Both underdoped cuprates and the \textit{n}=8 nickelates do not hold long range magnetic order, but the dispersiveless and narrow magnetic excitations in the \textit{n}=8 r-RP contrast with the overdamped dispersive high energy paramagnons in underdoped cuprates \cite{Le_Tacon_2011,Dean_2013,Guarise_2014}. Furthermore, the strength of the spin excitations in the reduced RP nickelates seems to be uncorrelated with the number of NiO$_2$ layers, unlike the RP cuprates \cite{Wang_2022,Dean_2014}, presumably highlighting the key role of chemical disorder upon reduction or reoxidation \cite{Hu_2024,Ranna_2025,Ortiz_2025,Pan2026}, frustrated longer-range hoppings, or competing ordered states.

To conclude, we have disentangled the low energy excitations in the \textit{n}= 8 parent and reduced Ruddlesden-Popper nickelates. Notably, the p-RP develops a long-range spin density wave with propagation vector q$_\mathrm{SDW}$=($\frac{1}{4}\ , \frac{1}{4}$), analog to the bilayer RP. Long-range order is suppressed upon oxygen reduction, while the RIXS spectra resolves a small enhancement of the in-plane magnetic interactions in the r-RP, as compared with the non-superconducting p-RP. Although at this stage we cannot directly link the emergence of superconducting correlations in the r-RP to the hardening of the magnetic excitations, our results highlight the critical role of the \textit{p}$_z$ orbital of the O$_\mathrm{ap}$ to tune the collective ground states in the \textit{n}=8 nickelates. 
 
\section*{ACKNOWLEDGEMENTS}  
We acknowledge Mark Dean for fruitful discussions and critical reading of the manuscript. We thank the European Synchrotron Radiation Facility, France, for time on beamline ID32 for polarimetric RIXS under proposal HC-5628. This work was supported by the MINECO of Spain, projects PID2021-122609NB-C21 and PID2024-161503NB-C21. A.K. and S.B-C. acknowledge financial support by the European Union Next Generation EU/PRTR-C17.I1, as well as by IKUR Strategy under the collaboration agreement between IKERBASQUE Foundation and DIPC on behalf of the Department of Education of the Basque Government. C.L. was supported by the European Research Council (ERC) under the European Union’s Horizon 2020 research and innovation program (Grant Agreement No. 101020833). A.S.B. acknowledges NSF Grant No. DMR-2323971. G.A.P., D.F.S., and J.A.M are supported by U.S. Department of Energy (DOE), Office of Basic Energy Sciences, Division of Materials Sciences and Engineering, under Award No. DE-SC0021925.  G.A.P. and D.F.S. acknowledge additional support under NSF Graduate Research Fellowship Grant No. DGE-1745303. The Flatiron Institute is a division of the Simons Foundation. E.H.d.S.N. acknowledges support from the National Science Foundation under grant number DMR-2034345. Work at UC San Diego was supported by the National Science Foundation under Grant No. DMR-2145080. F.B. acknowledges support from the Natural Sciences and Engineering Research Council of Canada, the Canada Research Chairs Program, and the Fonds de recherche du Quebec - Nature et Technologies

\bibliography{apssamp}

\end{document}